\documentclass[preprint,letterpaper]{JHEP3}

\usepackage{epsfig}

\title{Effect of charged partons on black hole production at the Large
Hadron Collider}   

\author{Douglas M. Gingrich\\ 
Centre for Particle Physics, Department of Physics, University of
Alberta, Edmonton, AB T6G 2G7 Canada\\
TRIUMF, Vancouver, BC V6T 2A3 Canada\\ 
E-mail: \email{gingrich@ualberta.ca}}

\abstract{The cross section for black hole production in hadron colliders is
calculated using a factorization hypothesis in which the parton-level
process is integrated over the parton density functions of the protons.    
The mass, spin, charge, colour, and finite size of the partons are
usually ignored. 
We examine the effects of parton electric charge on black hole
production using the trapped-surface approach of general relativity. 
Accounting for electric charge of the partons could reduce the black
hole cross section  by one to four orders of magnitude at the Large
Hadron Collider.
The cross section results are sensitive to the Standard Model brane
thickness.
Lower limits on the amount of energy trapped behind the event horizon 
in the collision of charged particles are also calculated.
}


\received{December 9, 2006}
\accepted{February 1, 2007}

\begin{document}

\section{\label{sec1}Introduction}

Models of large~\cite{ADD1,ADD2,ADD3} or warped~\cite{RS1,RS2} extra
dimensions allow the fundamental scale of gravity to be as low as the
electroweak scale. 
For energies above the gravity scale, black holes can be produced in
particle collisions.
This opens up the possibility to produce black holes at the Large Hadron
Collider (LHC).
Once formed, the black hole will decay by emitting Hawking
radiation~\cite{Hawking1}. 
The final fate of the black hole is unknown since quantum gravity will
become important as the black hole mass approaches the Planck scale.
If black holes are produced at the LHC, detecting them will not only
test general relativity and probe extra dimensions, but will also teach
us about quantum gravity.

Early discussions of black hole production in colliders postulated a
$\pi r_\mathrm{h}^2$ form for the cross section, where $r_\mathrm{h}$ is
the horizon radius of the black hole formed in the parton scattering
process~\cite{Banks,Giddings1,Dimopoulos1}.    
Calculations based on classical general relativity have had limited
success in improving the cross section
estimates~\cite{Eardley,Yoshino1}. 
The effects of mass, spin, charge, colour, and finite size of the
incoming particles are usually neglected in these calculations.
The effects of finite size have been
examined~\cite{Kohlprath,Giddings2} and only recently have angular
momentum~\cite{Yoshino2} or charge been discussed~\cite{Mann}. 
Although these results are far from complete, they do indicated that the 
simple geometric cross section is correct if multiplied by a formation
factor of order unity~\cite{Gingrich}.

General relativistic calculations of the cross section have usually 
been performed using the trapped-surface approach.
The two incoming partons are modeled as Aichelburg-Sexl shock
waves~\cite{Aichelburg}.  
Spacetime is flat in all regions of space except at the shocks.
The union of these shock waves defines a closed trapped surface.
Black hole formation can be predicted by identifying a future trapped
surface, with no need to calculate the gravitational field. 

The trapped-surface approach was first applied to TeV-scale gravity
calculations by Eardley and Giddings~\cite{Eardley} in four dimensions. 
Thier work was extended to the $D$-dimensional case numerically by
Yoshino and Nambu~\cite{Yoshino1}. 
The numerical studies were improved by Yoshino and
Rychkov~\cite{Yoshino2} by analyzing the closed trapped surface on a 
different slice of spacetime.
These general relativistic calculations have enabled lower limits to be
obtained for the black hole production cross section of colliding
particles in TeV-scale gravity scenarios.

Since black holes are highly massive objects, the momentum fraction of
the partons in the protons that form them must be high.
Thus typically valence quarks will be involved in black hole formation.
This means the most probable charge of the black hole in proton
collisions will be $+4/3$.
Since the gravitational field of each particle is determined by its
energy-momentum tensor, charge should affect the black hole formation.
First exploratory work by Yoshino and Mann~\cite{Mann} obtained a
condition on the electric charges of the colliding particles for a
closed trapped surface to form.  
The results depend on the Standard Model brane thickness.
Since the LHC is about to start up, it is useful to investigate ideas,
such as these, that modify black hole production.

In this paper, we use the Yoshino and Mann charge condition in its
general form and build on their work by examining the effect of charge
on black hole production at the LHC.
The cross section is obtained by summing over all possible parton pairs
in the protons.
By using the parton density functions of the proton and applying the
charge condition, we obtain the black hole cross section. 
The amount of available energy that goes into the black hole formation
is also examined.

An outline of this paper is as follows.
We first review the trapped-surface approach in section~\ref{sec2}.
Then in section~\ref{sec3}, we examine the Reissner-Nordstr{\"o}m
spacetime of charged particles in higher dimensions.
Following Yoshino and Mann~\cite{Mann}, we obtain the condition for an
apparent horizon in section~\ref{sec4}. 
Since electric charge is confined to the three-brane, we relate the
higher-dimensional charge in the Reissner-Nordstr{\"o}m metric to the
Standard Model electric charge as the second step in our approach.
We then translate the condition for apparent horizon formation into a
condition for charged partons to form a black hole at the LHC.   
Lower limits on the amount of energy trapped behind the apparent horizon
are shown in section~\ref{sec5}.
In section~\ref{sec6}, we apply the charge condition to the calculation
of the black hole cross section for different values of the Planck
scale, number of dimensions, and Standard Model brane thickness.
We conclude with a discussion in section~\ref{sec7}.

\section{\label{sec2}Trapped surfaces and apparent horizon}

In this section, we review the concepts of Aichelburg-Sexl shock waves,
trapped surfaces, and the apparent horizon.
A charged particle can be modelled using the Reissner-Nortsr{\"o}m
metric. 
By boosting the Reissner-Nortstr{\"o}m spacetime and taking the lightlike
limit, we obtain an approximation of a charged ultrarelativistic
particle.  
The gravitational field of the point massless particle is thus
infinitely Lorentz contracted and forms a longitudinal plane-fronted
Aichelburg-Sexl gravitational shock wave.  
Except at the shock wave, spacetime is flat before the collision.
By combining two Aichelburg-Sexl shock waves, we set up a head-on
collision of ultrelativistic particles.
At the instance of collision, the two shock waves pass through one
another, and interact nonlinearly by shearing and focusing.
After the collision, the two shocks continue to interact nonlinearly
with each other and spacetime within the future lightcone of the
collision becomes highly curved. 

To describe this nontrivial collision process, we choose the lightcone
coordinates $u=t-z$ and $v=t+z$. 
Four regions of spacetime can be identified.

\begin{description}
\item[\rm Region\quad I:] $u<0, v<0$, before the collision, 
\item[\rm Region\ \ II:]  $u>0, v<0$, after the wave at $u=0$ has passed, 
\item[\rm Region III:]    $v>0, u<0$, after the wave at $v=0$ has passed, 
\item[\rm Region  IV:]    $u>0, v>0$, interaction region after the waves
                                      have passed.
\end{description}

\noindent
Except at the shock waves, spacetime is flat in regions I, II, and III
before the collision.  
No one has been able to calculate the metric in the future of the
collision (non-linear region IV) except perturbatively in the distance
far from the interaction $u=v=0$~\cite{Eath1,Eath2,Eath3}.   
It is possible to investigate the collision on the slice $u \le
0, v = 0$ and $v \le 0, u=0$.
It is also possible to proceed with the analysis on the slice of the
future lightcone of the shock collision plane, given by the union of
the outgoing shocks $u=0, v\ge 0$ and $v=0, u\ge 0$. 
This is the future most slice that can be used without knowledge of
region IV.

The different regions of spacetime can be examined for trapped surfaces.  
We search for marginally trapped-surface formation on a slice of
spacetime $u=0$, $v\ge 0$ and $v=0$, $u\ge 0$. 
A marginally trapped surface is define as a closed spacelike
($D-2$)-surface, the outer null normals of which have zero
convergence~\cite{Ellis}.   
Moving a small distance inside the marginally trapped surface
one can find a true closed trapped surface with negative convergence.
In physical terms this means that there is a closed surface whose normal
null geodesics do not diverge, and so are trapped by gravity.
For a Schwarzschild black hole, the marginally trapped surface is a
sphere around the singularity, which happens to coincide with the
event horizon~\cite{Kang}.

An apparent horizon is the outermost marginally trapped surface.
The existence of a marginally trapped surface means either that the
marginally trapped surface is the apparent horizon, or that an apparent
horizon exits in the exterior of the marginally trapped surface.  
Existence of an apparent horizon implies the presence of a singularity
in the future. 
Assuming cosmic censorship~\cite{Penrose}, this singularity must be
hidden behind an event horizon, and we may conclude that a black hole
will form. 
Moreover, the black hole horizon must lie outside the closed trapped
surface. 
Formation of the apparent horizon is then a sufficient condition for
formation of a black hole for which the event horizon is outside the
apparent horizon~\cite{Wald}. 
Thus if one can prove the existence of a trapped surface, then one knows 
that in the future the solution will involve a black hole.

The area of a marginally trapped surface is a lower bound on the area of 
the apparent horizon. 
Using this information, one can estimate the event horizon area and,
via the area theorem~\cite{Hawking2}, the mass of the formed black
hole. 
Since the black hole horizon is always in the exterior region, the
trapped surface method gives only a lower bound on the final black hole
mass. 
The black hole mass can have any value between this bound and the centre
of mass energy of the collision.

\section{\label{sec3}\boldmath{$D$}-Dimensional Reissner-Nordstr{\"o}m
spacetime} 

The Reissner-Nordstr{\"o}m solution describing the gravitation field of a
point particle of mass $m$ and electric charge $q$ in $D$ dimensions in
spherical coordinates ($T,R,\Phi_1,\ldots,\Phi_{D-2}$) is  

\begin{equation}
ds^2 = -g(R)dT + g(R)^{-1}dR^2 + R^2d\Omega^2_{D-2} \, ,
\end{equation}

\noindent
where

\begin{eqnarray}
g(R) & = & 1 - \frac{2M}{R^{D-3}} + \frac{Q^2}{R^{2(D-3)}} \, ,
\label{eq2} \\
M & = & \frac{8\pi G_Dm}{(D-2)\Omega_{D-2}} \, , \label{eq3} \\ 
Q^2 & = & \frac{8\pi G_Dq^2}{(D-2)(D-3)} \, , \label{eq4}
\end{eqnarray}

\noindent
$G_D$ is the $D$-dimensional gravitational constant, and $d\Omega_{D-2}$
and $\Omega_{D-2}$ are the line element and volume of a
($D-2$)-dimensional unit sphere, given by

\begin{equation} \label{eq5}
\Omega_{D-2} = \frac{2\pi^{(D-1)/2}}{\Gamma[(D-1)/2]} \, ,
\end{equation} 

\noindent
where $\Gamma$ is Euler's Gamma function.
The energy-momentum tensor used in Einstein's equation is that of the
electromagnetic field in the spacetime that results from the charge on
the particle. 
The metric is a unique spherically symmetric asymptotically flat
solution of the Einstein-Maxwell equations and is locally similar
to the Schwarzschild solution. 
The Reissner-Nordstr{\"o}m solution does not describe the spin, magnetic
moment, or colour charge of a particle. 

The condition for the existence of an event horizon in $D$-dimensional
Reissner-Nordstr{\"o}m spacetime ($ Q^2 \le M^2$) is 

\begin{equation} \label{eq6}
|q| \le \frac{m}{\Omega_{D-2}} \sqrt{\frac{8\pi G_D(D-3)}{D-2}} \, .
\end{equation}

\noindent
We will return to this condition after we have related the electric
charge in higher-dimensional Maxwell theory to the electric charge
in four dimensions. 
For the moment, we are ignoring that the electric charge of the
particle is confined to the Standard Model three-brane.
We will consider the black hole production process as happening
in flat $D$-dimensional spacetime since the horizon radius is much small
than the compactification radius.  

Before Lorentz boosting the Reissner-Nordstr{\"o}m metric, it is
convenient to convert the metric to isotropic coordinates
($\bar{T},\bar{Z},\bar{R},\bar{\Phi}_1,\ldots,\bar{\Phi}_{D-3}$), where 
$\bar{T} = T$, $\bar{Z} = Z$, $\bar{R} = \sqrt{\bar{Z}^2 +\bar{r}^2}$,
$R^2d\Omega_{D-2}^2 =d\bar{Z}^2 + d\bar{r}^2 + \bar{r}^2
d\bar{\Omega}_{D-3}^2$, and  

\begin{equation}
R = \bar{R} \left( 1 + \frac{M}{\bar{R}^{D-3}} + \frac{M^2 -
Q^2}{4\bar{R}^{2(D-3)}} \right)^{\frac{1}{D-3}} .  
\end{equation}

When the Reissner-Nordstr{\"o}m metric is boosted, the rest mass $m$ and
charge $q$ are fixed and boosted to a finite value of $\gamma$.
In the ultrarelativistic limit ($\gamma \to \infty$)
both terms in eq.~(\ref{eq2}) diverge unless we take both $m$ and $q$ to
vanish in this limit.   
Thus, we boost the Reissner-Nordstr{\"o}m solution by taking the limit
of large boost and small $m$ with fixed total energy

\begin{equation}
E = \gamma m \, ,
\end{equation}

\noindent
and small $q^2$ with fixed quantity

\begin{equation}
p_e^2 = \gamma q^2 \, .
\end{equation}

\noindent
This is consistent with the lightlike limit of a particle with mass and
electric charge.
Choosing the particle to move in the $+Z$ direction in $D$-dimensional
spacetime, the boosted coordinates are
($t,z,r,\phi_1,\ldots,\phi_{D-3}$), where $\bar{T} = \gamma (t -\beta
z)$, $\bar{Z} = \gamma (z -\beta t)$, $r = \bar{R}$, and $\phi_i =
\bar{\Phi}_i$. 
After the transformation, we take the lightlike limit.

Next we define the retarded and advanced times ($\bar{u} = t -z$ and
$\bar{v} = t +z$, also $\bar{r} = r$ and $\bar{\phi}_i = \phi_i$) to
obtain the coordinates
($\bar{u},\bar{v},\bar{r},\bar{\phi}_1,\ldots,\bar{\phi}_{D-3}$).    
This yields a finite result that is the charged version of the
$D$-dimensional Achelburg-Sexl metric~\cite{Mann,Lousto,Ortaggio}:

\begin{equation} \label{eq8}
ds^2 = -d\bar{u}d\bar{v} + d\bar{r}^2 + \bar{r}^2d\bar{\Omega}_{D-3} +
\Phi(\bar{r}) \delta(\bar{u}) d\bar{u}^2 \, , 
\end{equation}

\noindent
where

\begin{equation}
\Phi(\bar{r}) = \left\{ \begin{array}{ll} -8G_DE\ln\bar{r}
-\frac{2a}{\bar{r}} & (D = 4) \, , \\ \nonumber
\frac{16\pi G_D
E}{(D-4)\Omega_{D-3}\bar{r}^{D-4}} -\frac{2a}{(2D-7)\bar{r}^{2D-7}} &
(D \ge 5) \, ,
\end{array} \right. 
\end{equation}

\noindent
and 

\begin{equation} \label{eq10}
a = \frac{2\pi(4\pi G_D p_e^2)}{D-3} \frac{(2D-5)!!}{(2D-4)!!} \, .
\end{equation}

\noindent
The charge dependence is entirely contained in the general charge
parameter $a$.
The function $\Phi$ depends only on the transverse radius $\bar{r}
=  \sqrt{\bar{x}^i\bar{x}_i}$.
The Aichelburg-Sexl metric is a solution for a point particle (delta
function source) moving at the speed of light.
The metric eq.~(\ref{eq8})-(\ref{eq10}) reduces to the usual
Aichelbrg-Sexl metric in the limit $q \to 0$.
For the particles we will consider in this study, $\gamma \gtrsim
5\times 10^3$ and the mean value of $\gamma$ is about $7\times 10^5$.
The charged version of the Aichelberg-Sexl metric is thus a good
approximation to an ultrarelativistic massive charged particle with
finite, but large, $\gamma$. 

The delta function in eq.~(\ref{eq8}) indicates that the $(\bar{u},
\bar{v}, \bar{r}, \bar{\phi}_i)$ coordinates are discontinuous at
$\bar{u}=0$. 
These coordinates are unsuitable for analysing the behaviour of
geodesics crossing the shock at $\bar{u}<0$, which is necessary for
understanding the causal structure.
In the following, we define

\begin{equation}
r_0 = \left( \frac{8\pi G_D E}{\Omega_{D-3}} \right)^\frac{1}{D-3} \equiv 1
\end{equation}

\noindent
as the unit of length.
We introduce new coordinates ($u,v,r,\phi_i$), which are continuous and
smooth across the shock using the transformations   

\begin{eqnarray}
\bar{u} & = & u \, , \\
\bar{v} & = & v + F(u,r) \, , \\
\bar{r} & = & G(u,r) \, , \\
\bar{\phi}_i & = & \phi_i \, ,
\end{eqnarray}

\noindent
where $F(u,r) = 0$ and $G(u,r) = r$ for $u<0$.
In these coordinates, we require $v$, $r$, and $\phi_i$ equal a
constant to be a null geodesic with affine parameter $u$.
The metric in these coordinates becomes

\begin{equation}
ds^2 = -du dv + G_{,r}^2 dr^2 + G^2 d\Omega_{D-3}^2 \, ,
\end{equation}

\noindent
where $G$ and $G_{,r}$ are explicitly given by~\cite{Mann}

\begin{eqnarray}
G & = & r + \frac{u\theta(u)}{r^{D-3}} \left( 1 -\frac{a}{r^{D-3}}
\right) , \\ 
G_{,r} & = & 1 + (D-3) \frac{u\theta(u)}{r^{D-2}} \left( 1 -
\frac{2a}{r^{D-3}} \right) ,
\end{eqnarray}

\noindent
where $\theta(u)$ is the Heaviside step function.
Both geodesics and their tangents are now continuous across the shock,
and two coordinate singularities appear in the region $u>0$.
These singularities have been analyzed in Ref.~\cite{Mann}.

\section{\label{sec4}Condition for apparent horizon formation}

To setup the two-particle head-on collision, we consider a second
identical shock wave traveling along $\bar{v}=0$ in the $-Z$ direction. 
By causality, the two shock waves will not be able to influence each
other until the shocks collide.
This means that we can superimpose two of the above solutions to
give the exact geometry outside the future lightcone of the collision of
the two shocks. 
We assume without loss of generality that the two particles have the
same energy $E$ but different charge parameters $p_e^{(1)}$ and
$p_e^{(2)}$. 

In the remainder of this section, we follow Yoshino and Mann~\cite{Mann}
directly in studying the apparent horizon on the slice $u>0$, $v=0$ and
$v>0$, $u=0$.  
The closed trapped surfaces are symmetric under rotation of the
transverse directions and the reflection $z\to -z$.
Because the system is axisymmetric, the location on the apparent
horizon surface on each side of $z$ is given by a function of
$r$. 
We assume the apparent horizon is given by the union of two surfaces
$S_1$ and $S_2$, where

\begin{eqnarray}
S_1: u = h^{(1)}(r) & \quad & (r_\mathrm{min} \le r \le
r_\mathrm{max}^{(1)}) \quad \mathrm{on} \quad u\ge 0\ \mathrm{and}\ v=0 , \\ 
S_2: v = h^{(2)}(r) & \quad & (r_\mathrm{min} \le r \le
r_\mathrm{max}^{(2)}) \quad \mathrm{on} \quad v\ge 0\ \mathrm{and}\ u=0 ,  
\end{eqnarray}

\noindent
where $h^{(1)}$ and $h^{(2)}$ are monotonically increasing functions of
$r$. 
When $S_1$ and $S_2$ cross $u=v=0$, $r=r_\mathrm{min}$.
Continuity of the metric at the apparent horizon requires $S_1$ and
$S_2$ to coincide with each other at $u=v=0$.
At $r_\mathrm{max}^{(1)}$ and $r_\mathrm{max}^{(2)}$, we require 
$h^{(1)}(r)$ and $h^{(2)}(r)$ to cross the coordinate singularity. 
The surface becomes a closed trapped surface by the above arguments.

Following Yoshino and Mann~\cite{Mann}, one derives the equations for
$h^{(1)}(r)$ and $h^{(2)}(r)$, and the differential equation for
the apparent horizon is then obtained. 
The boundary condition that must be imposed at $r=r_\mathrm{min}$ is

\begin{equation}
h_{,r}^{(1)}(r_\mathrm{min}) h_{,r}^{(2)}(r_\mathrm{min}) = 4 \, ,
\end{equation}

\noindent
where both $h_{,r}^{(1)}(r_\mathrm{min})$ and
$h_{,r}^{(2)}(r_\mathrm{min})$ are positive. 
Using this boundary condition, the apparent horizon equation is solved
and the boundary condition become 

\begin{equation} \label{eq23}
x^4 = (x-a_1)(x-a_2) \, ,
\end{equation}

\noindent
where $x \equiv r_\mathrm{min}^{D-3}$.
This equation determines the value of $r_\mathrm{min}$.
The apparent horizon exists if, and only if, there is a solution to
eq.~(\ref{eq23}) with $x > a_1$ and $x > a_2$. 

\subsection{Condition on general charges}

The special cases of collisions of particles with the same charge, and
collisions of a charged and a neutral particle have been previously
examined~\cite{Mann}.
We examine the case of general charge parameters and note the simplifying
cases. 

Equation~(\ref{eq23}) has four roots.
For $a_1>0$ and $a_2>0$ two solutions will have $x<a_1$ and $x<a_2$, and not
correspond to an apparent horizon.
We investigate the other two solutions.
Figure~\ref{quartic} shows two representative cases of eq.~(\ref{eq23}).
We see that an apparent horizon exists if, and only if, the local minimum
in the $x>0$ region is less than or equal to zero. 
The location of the minium is given by differentiating the quartic 
equation, eq.~(\ref{eq23}), and solving the resulting cubic equation for  
the positive root. 
The solution is

\FIGURE{
\epsfig{file=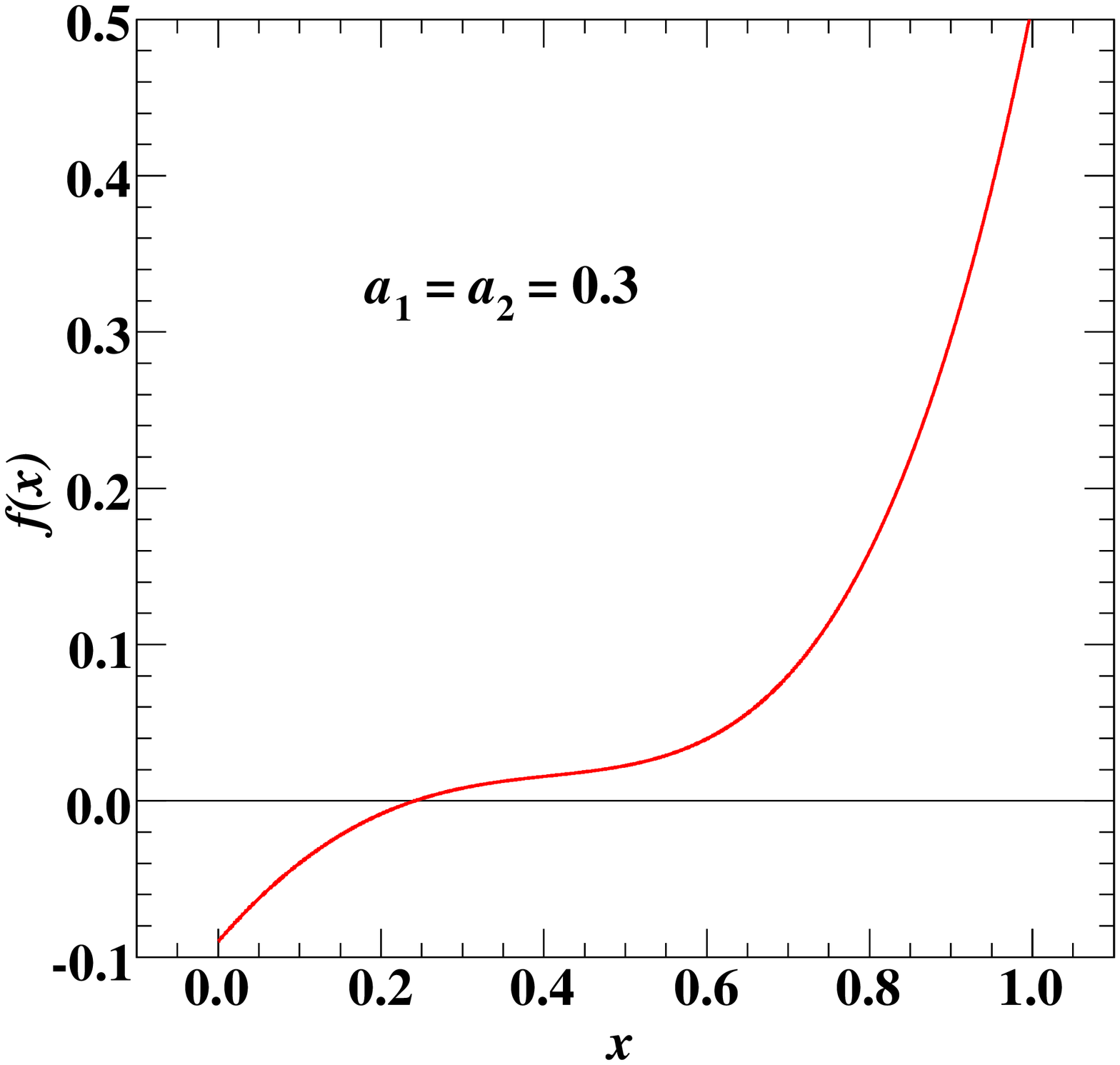,width=7.5cm}
\epsfig{file=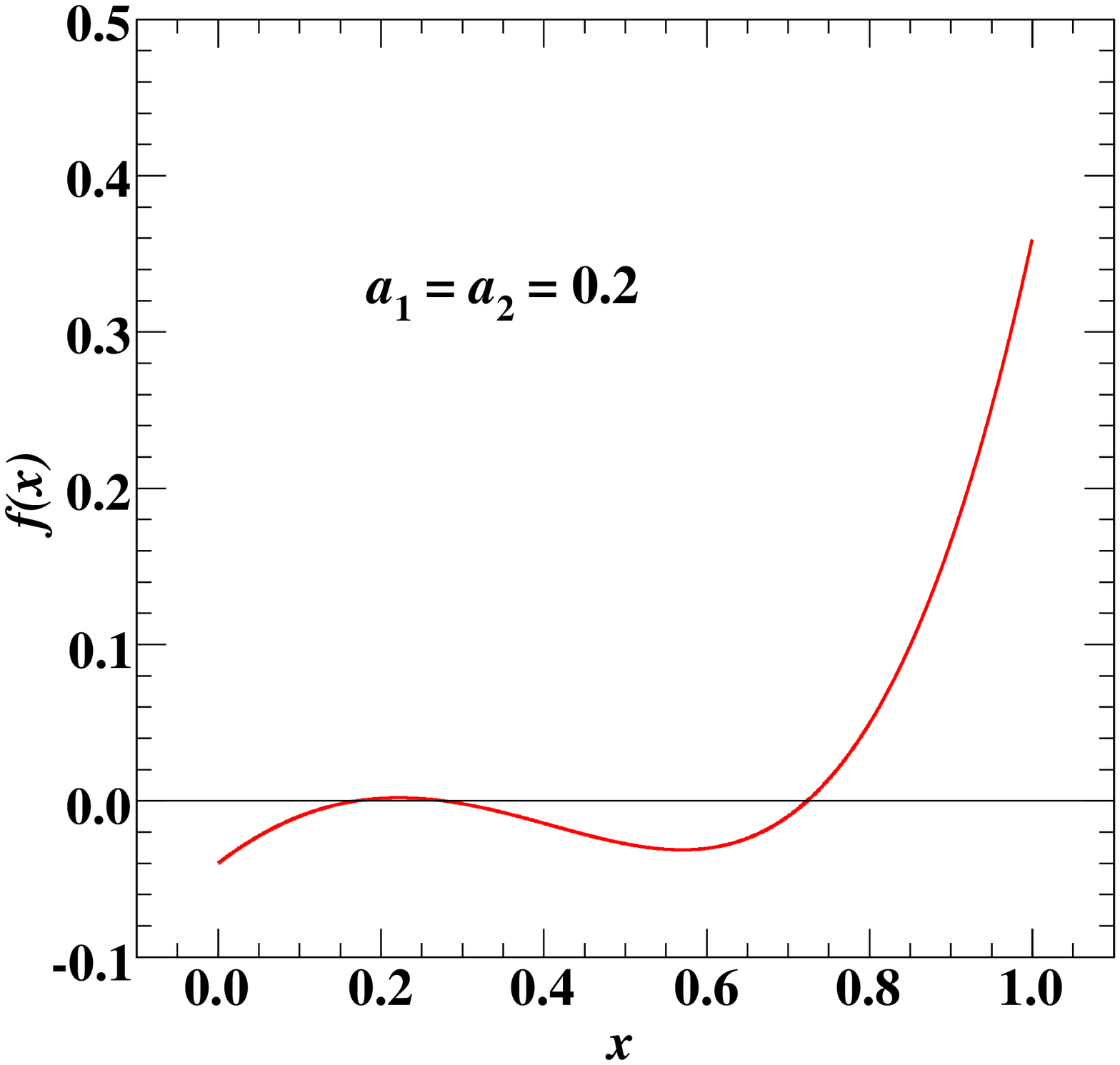,width=7.5cm}
\caption{Representative cases for the quartic equation $f(x) = x^4
-(x-a_1)(x-a_2)$.} 
\label{quartic}
}

\begin{equation}
x = \sqrt{\frac{2}{3}} \cos \left( \frac{\pi-\theta}{3} \right) \, ,
\end{equation}

\noindent
where

\begin{equation}
\tan\theta = \sqrt{\left( \frac{2}{3} \right)^3 \frac{1}{(a_1+a_2)^2}
-1} \, .
\end{equation}

\noindent
Substituting this value into eq.~(\ref{eq23}) and drawing the contour,
we find the region for apparent horizon formation in the
$(a_1,a_2)$-plane, as shown in fig.~\ref{ah}.  
We see that both $a_1$ and $a_2$ must be sufficiently small for apparent
horizon formation. 
For two particles of equal charge: $a_1 = a_2 = 1/4$ gives $x=1/2$,
which is a solution of eq.~(\ref{eq23}).
For one charged particle and one neutral particle: $a_1 = a
=2/(3\sqrt{3})$ and $a_2 = 0$ gives $x=1/\sqrt{3}$, which is also a
solution of eq.~(\ref{eq23}). 

\FIGURE{
\epsfig{file=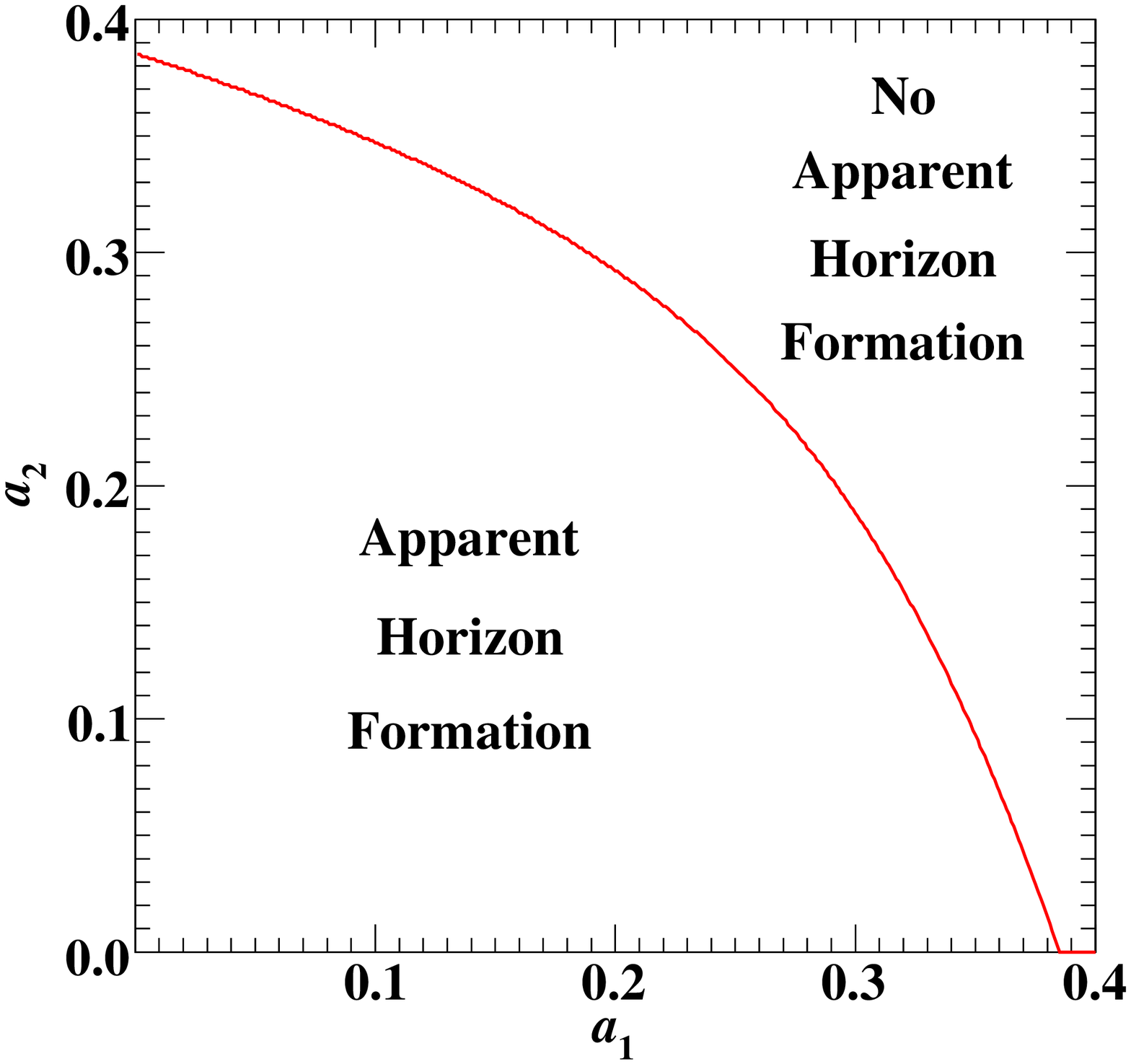,width=9cm}
\caption{Region of the apparent horizon formation in the
$(a_1,a_2)$-plane.} 
\label{ah}
}

We can understand the requirement on $a_1$ and $a_2$ physically as
follows. 
Since $a_1$ and $a_2$ are proportional to $(p_e^{(1)})^2$ and
$(p_e^{(2)})^2$, the condition derived in eq.~(\ref{eq23}) does not
depend on the sign of the charge of either particle.
This is because the gravitational field due to each charge is generated
by an electromagnetic energy-momentum tensor $T_{\mu\nu}^{(\mathrm{em})}
\sim p_e^2\delta(\bar{u})/\bar{r}^{2D-5}$ that depends on the square of the
charge. 
The gravitational field induced by  $T_{\mu\nu}^{(\mathrm{em})}$ of the
incoming particles is repulsive, and its affect becomes dominant around
the centre. 
As the value of $a$ increases, the repulsive region becomes larger,
preventing formation of the apparent horizon.
The critical value of $a$ for apparent horizon formation occurs when the 
repulsive gravitational force due to the electric field becomes
equivalent to the self-attractive force due to the energy of the system.

\subsection{Condition on parton electric charges}

The approach for handling the confinement of the electric field
to the Standard Model three-brane is far from clear.
So far, we have ignored this effect by using the $D$--dimensional
Einstein-Maxwell theory.
Assuming the boosted Reissner-Nordstr{\"o}m metric represents the
gravitational field of an elementary particle with electric charge
moving at high speed, we develop the relationship between the electric
charge in four dimensions $q_4$ and the charge in higher-dimensional
Maxwell theory $q$.   
For two particles in $D$ dimensions with the same charge at rest, for
example, the force between them is 

\begin{equation}
F = \frac{q^2}{r^{D-2}} \, .
\end{equation}

\noindent
Since we have been using Gaussian units, the factor of $1/\Omega_{D-2}$
($1/4\pi$ in the case of four dimensions) is absorbed into the
definition of the charge. 
If the gauge fields are confined to the Standard Model brane, the only
characteristic length scale is the width of the brane, which should be
of the order of the Planck length.
We introduce the constant $C_\mathrm{brane}$:

\begin{equation}
\frac{1}{M_D} \to \frac{C_\mathrm{brane}}{M_D} \, ,
\end{equation}

\noindent
where $C_\mathrm{brane}$ is a dimensionless quantity of order unity.
For sufficiently large $r$, 

\begin{equation}
F \to \frac{q^2}{r^2} \left( \frac{M_D}{C_\mathrm{brane}} \right)^{D-4}=
\frac{q_4^2}{r^2} \, .
\end{equation}

\noindent
This condition is reasonable since the Compton wavelength $1/M_D$ of the
black hole is much smaller than its horizon radius $r_\mathrm{h}$.
Thus

\begin{equation} \label{eq30}
q^2 = q_4^2 \left( \frac{C_\mathrm{brane}}{M_D} \right)^{D-4} \, .
\end{equation}

\noindent
The brane thickness is a measure of how confined the Standard Model
electric charge is to the brane.
If the brane is thick, the Maxwell theory would be higher dimensional in
the neighbourhood of the particle.
We let

\begin{equation} \label{eq31}
q_4^2 = C_q^2 \alpha \, ,
\end{equation}

\noindent
where $C_q$ is the charge in units of elementary charge $e$ ($-1/3$ or
$+2/3$ for quarks and 0 for gluons) and $\alpha$ is the fine structure
constant.  
Our treatment of the electric charge has not fully taken into account
the effects of confinement of the electric field on the brane.
We have also ignored the brane tension and the structure of the extra
dimensions. 

Using eq.~(\ref{eq30}), eq.~(\ref{eq31}), 

\begin{equation}
G_D^{-1} = \frac{8\pi}{(2\pi)^{D-4}} M_D^{D-2} \, ,
\end{equation}

\noindent
and recalling the definition of the volume of the ($D-2$)-dimensional
unit sphere given by eq.~(\ref{eq5}), we obtain for eq.~(\ref{eq10}) 

\begin{equation} \label{eq32}
\frac{a}{r_0^{2(D-3)}} = C_q^2 \alpha \left( \frac{M_D}{m} \right)
\left( \frac{M_D}{E} \right) \pi 
\frac{\Omega_{D-3}^2}{D-3} \frac{(2D-5)!!}{(2D-4)!!} \left( 
\frac{C_\mathrm{brane}}{2\pi} \right)^{D-4} \, ,
\end{equation}

\noindent
where we have reintroduced the unit length $r_0$.
Choosing values for $\alpha$ and $m$, we can use eq.~(\ref{eq32}) to
study the condition for apparent horizon formation as a function of $D$,
$M_D$, and $C_\mathrm{brane}$.   
An apparent horizon will not occur at the instance of collision if the
brane is thick or if the spacetime dimensionality is low. 
Charge effects will not be significant at high energies.

\section{\label{sec5}Trapped energy}

We now consider how much energy is trapped behind the horizon in black
hole formation from charged-particle collisions.
From section~\ref{sec2}, we saw that the event horizon must be outside
the apparent horizon.  
The area theorem~\cite{Hawking2} states that the event horizon area can
never decreases.  
Hence, we naturally expect the apparent horizon mass to be defined by

\begin{equation}
M_\mathrm{AH} = \frac{(D-2)\Omega_{D-2}}{16\pi G_D} \left(
\frac{A_{D-2}}{\Omega_{D-2}} \right)^{\frac{D-3}{D-2}} \, ,
\end{equation}

\noindent
where $A_{D-2}$ is the ($D-2$)-dimensional area of the apparent
horizon give by

\begin{equation}
A_{D-2} = \frac{2}{D-2} \Omega_{D-3} r_0^{D-2} x^\frac{D-2}{D-3} \, . 
\end{equation}

\noindent
This mass provides a lower bound on the mass of the final black hole,
and thus $M_\mathrm{AH}$ is an indicator of the energy trapped behind
the event horizon. 
The parameter $x$ can be considered to be a function of $a_1$ and $a_2$,
with $x=1$ for $a_1=a_2=0$.
The results of Eardley and Giddings~\cite{Eardley} are reproduced for the
case of two neutral particles.  

Figure~\ref{eff} shows the behaviour of $M_\mathrm{AH}/\sqrt{\hat{s}}$ as a
function of $a_1$ and $a_2$ for $D=4$, where $\sqrt{\hat{s}}$ is the
centre of mass energy of the collision.
We find that $M_\mathrm{AH}/\sqrt{\hat{s}}$ decreases slowly with
increasing $a$, but drops rapidly near the maximum value of $a$.
Figure~\ref{effsame} (particles of same charge) and fig.~\ref{effneut}
(one neutral particle) show the behaviour of
$M_\mathrm{AH}/\sqrt{\hat{s}}$ as a function of $a$ for different values
of the number of dimensions. 
The horizon mass decreases with increasing number of dimensions.
In the higher-dimensional spacetime, the amount of energy trapped behind
the horizon decreases because the gravitational field distributes in the
space of the extra dimensions and only a small portion of the total
energy of the system can contribute to the horizon formation.
The non-trapped energy will be radiated away quickly after the formation 
of the black hole.

\FIGURE{
\epsfig{file=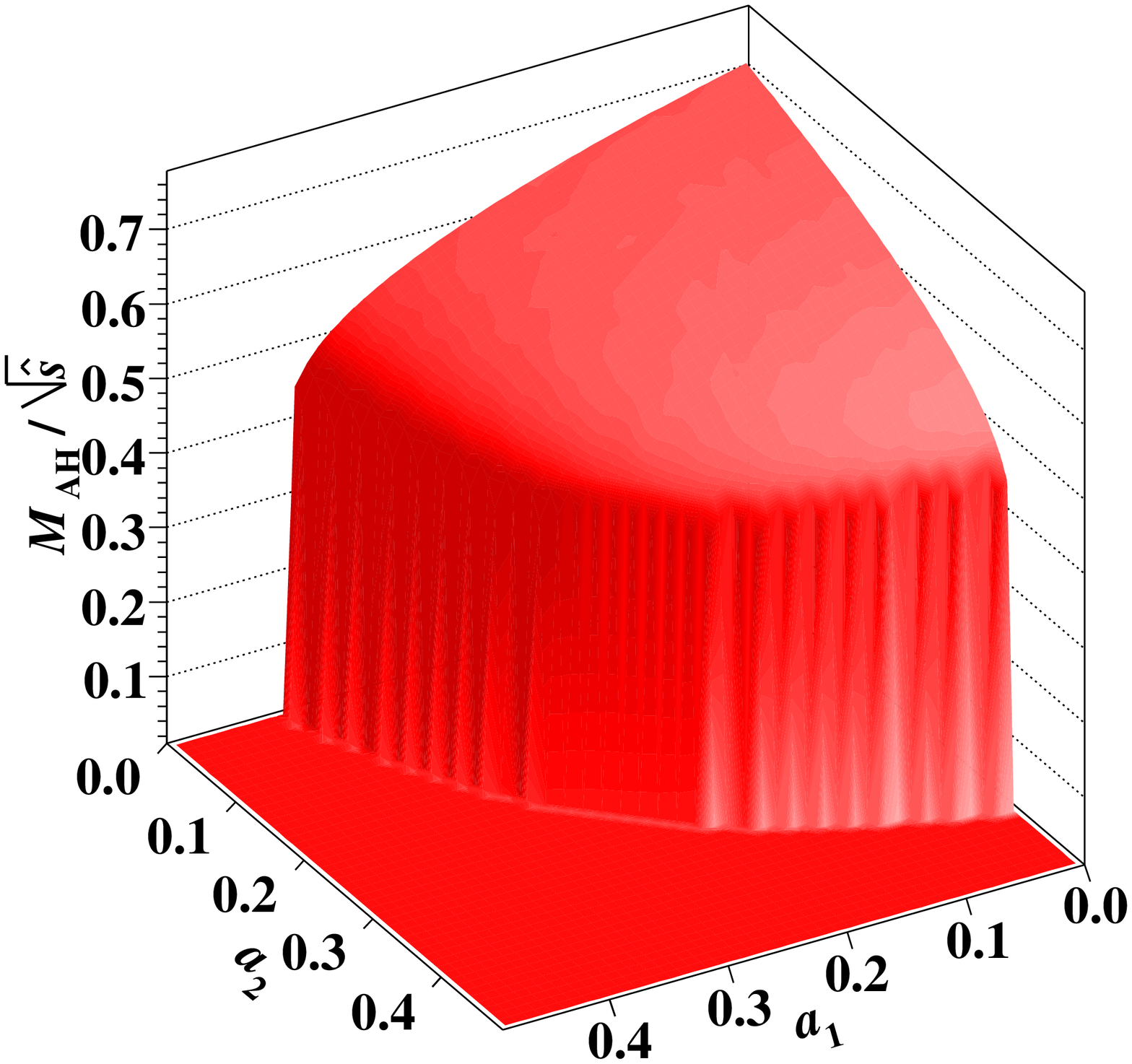,width=12cm}
\caption{Relation between the horizon mass
$M_\mathrm{AH}/\sqrt{\hat{s}}$ and the charge parameters $a_1$ and $a_2$
for $D=4$.} 
\label{eff}
}

\FIGURE{
\epsfig{file=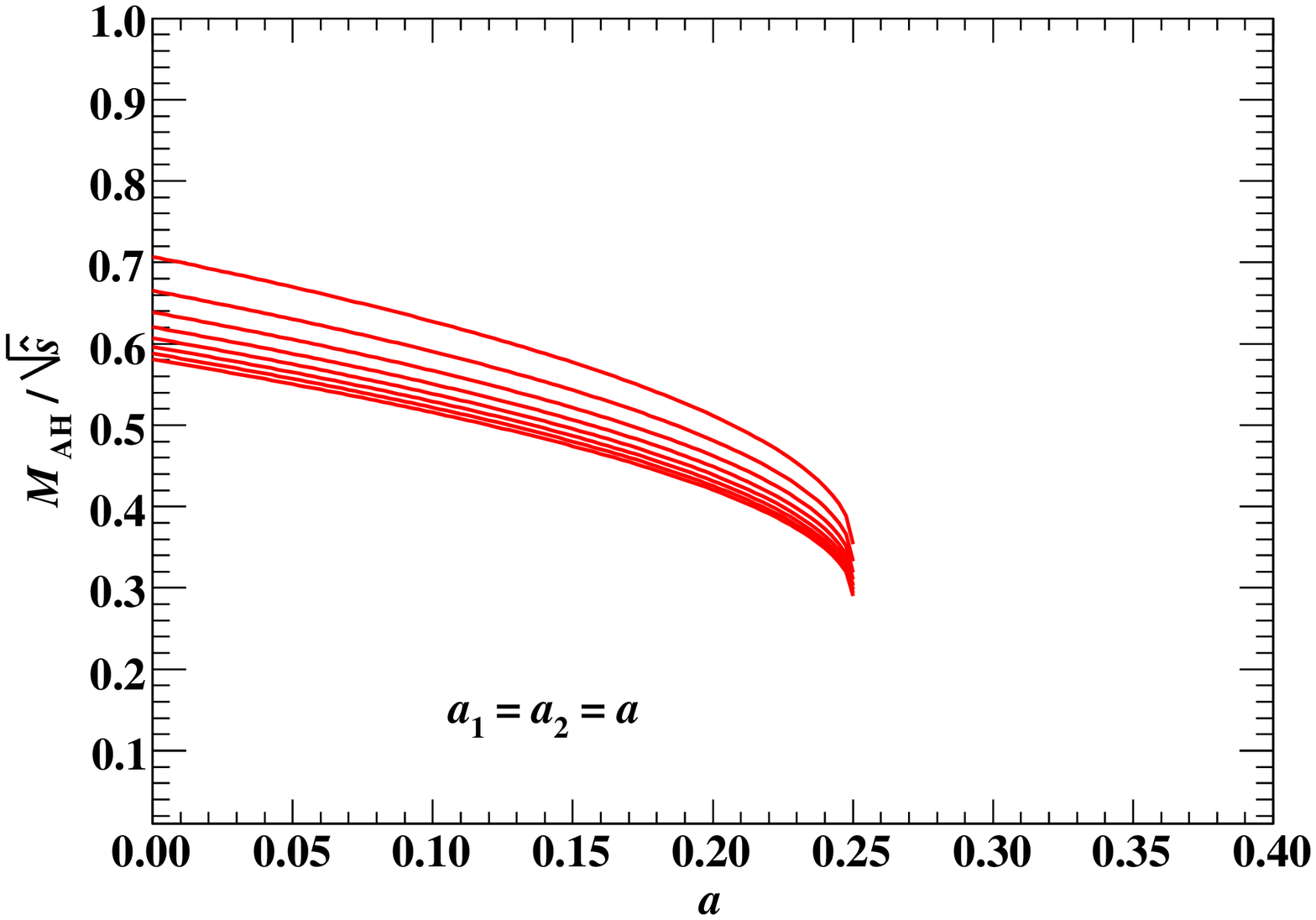,width=12cm}
\caption{Relation between the horizon mass $M_\mathrm{AH}/\sqrt{\hat{s}}$ and
the charge parameters $a_1=a_2=a$ (particles of equal charge) for
$D=4,\ldots,11$.} 
\label{effsame}
}

\FIGURE{
\epsfig{file=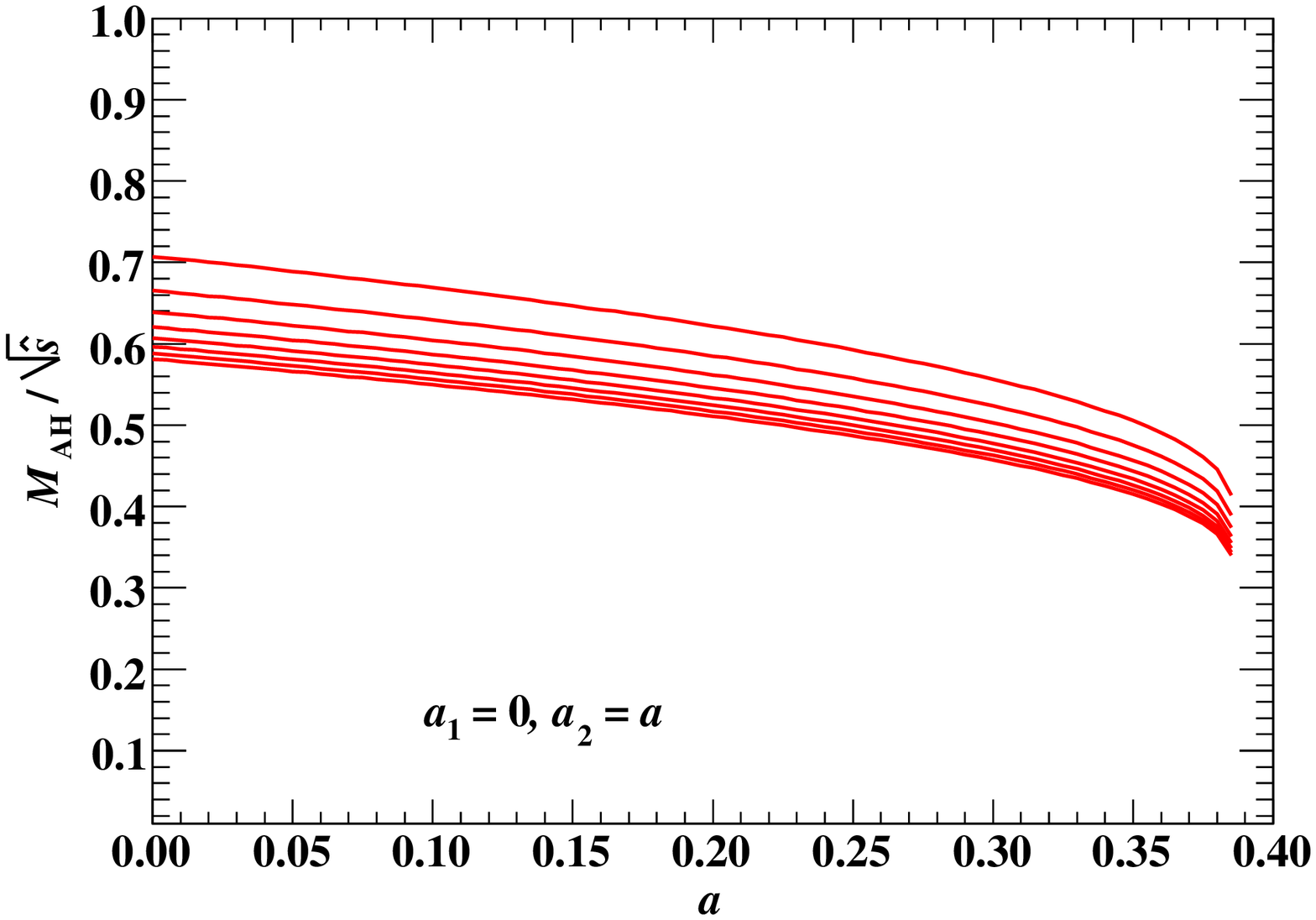,width=12cm}
\caption{Relation between the horizon mass $M_\mathrm{AH}/\sqrt{\hat{s}}$ and
the charge parameters $a_1=a$ and $a_2=0$ (one neutral particle) for
$D=4,\ldots,11$.} 
\label{effneut}
}

\section{\label{sec6}Effect of charged partons on the cross section}

The classical black hole cross section at the parton level is 

\begin{equation}
\hat{\sigma}_{ab\to\mathrm{BH}} = \pi r_\mathrm{h}^2 \, ,
\end{equation}

\noindent
where $r_\mathrm{h}$ depends on the mass of the black hole
$M_\mathrm{BH}$, the spacetime parameters $D$ and $M_D$, and $a$ and
$b$ are the parton types. 

Only a fraction of the total centre of mass energy $\sqrt{s}$ in a
proton-proton collision is available in the parton scattering process.
We define

\begin{equation}
s x_a x_b \equiv s\tau \equiv \hat{s} \, ,
\end{equation}

\noindent
where $x_a$ and $x_b$ are the fractional energies of the two partons
relative to the proton energies.
The total cross section can be obtained by convoluting the parton-level
cross section with the parton distribution functions (PDFs), integrating
over the phase space, and summing over the parton types.
Assuming all the available parton energy $\sqrt{\hat{s}}$ goes into
forming the black hole, the full particle-level cross section is 

\begin{equation} \label{eq35}
\sigma_{pp\to\mathrm{BH+X}}(M_\mathrm{BH}) = \sum_{a,b}
\int_{M_\mathrm{BH}^2/s}^1 dx_a \int_{M_\mathrm{BH}^2/(x_as)}^1 dx_b
f_a(x_a) f_b(x_b) \hat{\sigma}_{ab\to\mathrm{BH}}(\hat{s}=M_\mathrm{BH})
\, ,
\end{equation}

\noindent
where $f_a$ and $f_b$ are PDFs for the proton.
The sum is over all possible quark and gluon pairings.

Throughout this paper we use the CTEQ6L1 (leading order with leading
order $\alpha_s$) parton distribution functions~\cite{Pumplin} within
the LHAPDF framework~\cite{LHAPDF}. 
The momentum scale for the PDFs is set equal to the black hole mass for
convenience. 

In terms of the parton luminosity (or parton flux), we write the
differential form of eq.~(\ref{eq35}) as

\begin{equation}
\frac{d\sigma_{pp\to\mathrm{BH+X}}}{dM_\mathrm{BH}} =
\frac{dL}{dM_\mathrm{BH}} \hat{\sigma}_{ab\to\mathrm{BH}} \, , 
\end{equation}

\noindent
where

\begin{equation} \label{eq36}
\frac{dL}{dM_\mathrm{BH}} = \frac{2M_\mathrm{BH}}{s} \sum_{a,b}
\int_{M_\mathrm{BH}^2/s}^1 \frac{dx}{x} f_a\left( \frac{\tau}{x} \right)
f_b(x) \, . 
\end{equation}

\noindent
The differential cross section thus factorizes.
It can be written as the product of the parton cross section times a
luminosity function.
The parton cross section $\hat{\sigma}_{ab\to \mathrm{BH}}$ is
independent of the parton types and depends only on the black hole mass,
Planck scale, and number of dimensions. 
The parton luminosity function contains all the information about the
partons.
Besides a dependence on the black hole mass, it is independent of the
characteristics of the higher-dimensional space, i.e.\ the Planck scale
and number of dimensions.
The dependence of the black hole mass occurs only in the proportionality
and the limit of integration.

The transition from the parton-level to the hadron-level cross section
is based on a factorization formula.
The validity of this formula for the energy region above the Planck
scale is unclear.
Even if factorization is valid, the extrapolation of the parton
distribution functions into this transplanckian region based on Standard
Model evolution from present energies is questionable, since the
evolution equations neglect gravity.

For a fixed proton-proton centre of mass energy, the parton luminosity
function can be pre-calculated to obtain a function depending only on a
single mass parameter.
Figure~\ref{lumi} shows the parton luminosity function versus black hole
mass for $\sqrt{s} = 14$~TeV for different partons in the sum of
eq.~(\ref{eq36}). 
The solid line is for all partons, include sea quarks and gluons.
The dashed line shows the luminosity with the sea quarks removed.
The doted line shows the luminosity with the sea quarks and gluons
removed, and the dash-dotted line is for only gluons.
Figure~\ref{lumi} indicates that to a good approximation, we can ignore
the contribution from the sea quarks at high black hole masses.
The gluon-only contribution is the lower bound on the luminosity
function when the charged quarks do not contribute to the cross section.  

\FIGURE{\epsfig{file=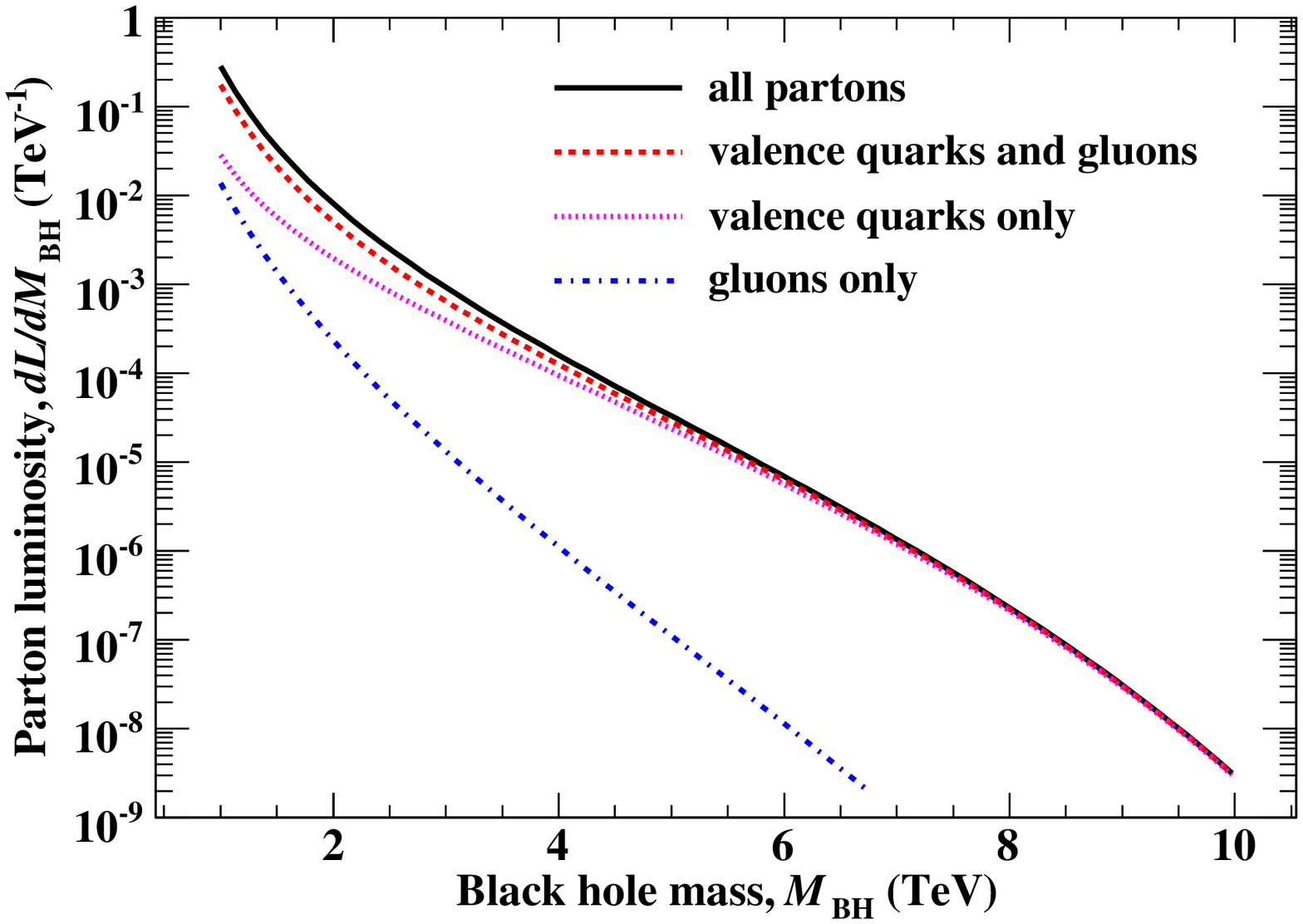,width=12cm}
\caption{Different contributions to the parton luminosity function
versus black hole mass.} 
\label{lumi}
}

Throughout the remainder of this analysis, we ignore the contribution to
the cross section from the sea quarks.
We work with parton luminosity, which is independent of $D$ and $M_D$.
Only the condition on which quarks to include in the sum of
eq.~(\ref{eq36}) depends on $D$ and $M_D$.
Thus the upper and lower bounds on the parton luminosity do not change
for different parameters. 
We take the running of the coupling constant into account; 
$\alpha$ ranges from about 1/126 to 1/122 over a black hole
mass range of 1~TeV to 10~TeV.
We choose $\alpha$ equal to 1/124 in the following calculations.
Because of the large momentum transfer in black hole production, we use
current quark masses. 
Quark masses of $m_\mathrm{d} =8$~MeV and $m_\mathrm{u} =4$~MeV are
chosen for the valence quarks in the proton.
To study eq.~(\ref{eq36}), we must first boost the partons to the
equal-energy frame to calculate eq.~(\ref{eq32}), and then determine if
the condition in fig.~\ref{ah} is satisfied. 
If it is, the parton pair is included in the sum in eq.~(\ref{eq36}).

Figure~\ref{adist} shows the distribution of values of the charges $a_1$
and $a_2$ for 11 dimensions, a Planck scale of 1.75~TeV, and a brane
thickness of 1. 
Distributions for same charges, different charges, and one neutral
parton combinations are clearly visible. 
The distributions fall off with increasing values of $a$.
The maximum values of $a_1$ and $a_2$ depend on the higher-dimensional
spacetime parameters $D$, $M_D$, and $C_\mathrm{brane}$.  
The bin representing gluon-gluon collisions $(a_1=a_2=0)$ is surrounded
by a region (not visible) of no events.
This vacated region increases with increasing Planck scale.

\FIGURE{\epsfig{file=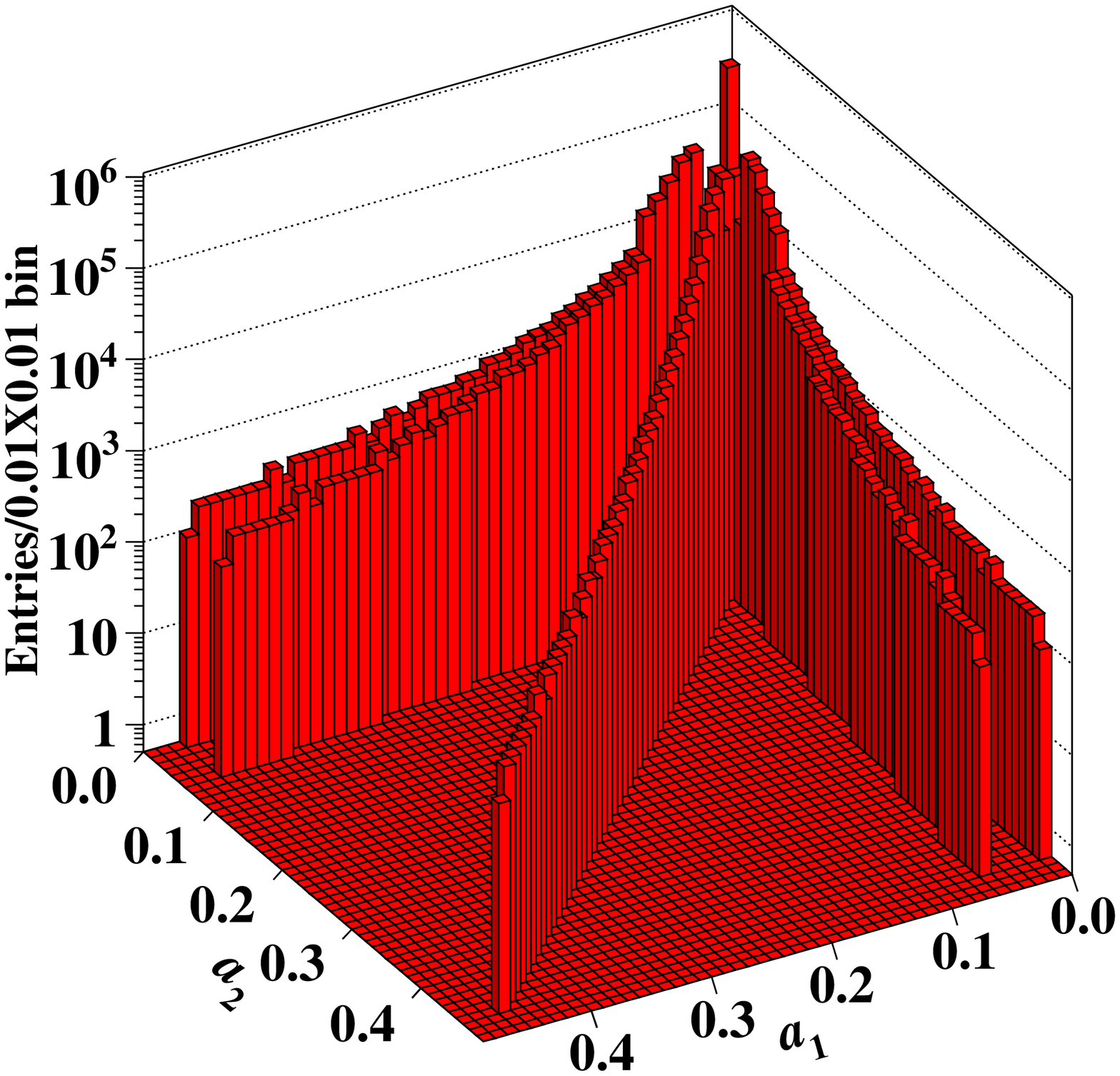,width=12cm}
\caption{Histrogram of values of the particle charges $a_1$ and $a_2$
in proton-proton collisions with $D=11$, $M_D=1.75$~TeV, and
$C_\mathrm{brane}=1$.} 
\label{adist}
}

Figure~\ref{charge} shows the parton luminosity for different brane
thicknesses for 11 dimensions and a Planck scale of 1~TeV. 
The top curve is the case when all the partons contribute to the cross
section, while the lower curve is the case when only the neutral gluons 
contribute to the cross section. 
The contributions of the different quarks in the intermediate region
depends on $M_\mathrm{BH}$, $D$, $M_D$, and $C_\mathrm{brane}$.
The thresholds for different quarks to contribute occur as a function of 
$M_\mathrm{BH}$ for fixed $D$, $M_D$, and $C_\mathrm{brane}$. 
The location of the thresholds may or may not occur in the mass region
of our plot. 
From fig.~\ref{charge}, we see that charge can affect any black hole mass
and the effect is very sensitive to the brane thickness.
The decrease in parton luminosity, and thus cross section, can range from
about one to four orders of magnitude over a black hole mass range of
$1-10$~TeV due to charge effects. 
The cross section is nontrivial only over a range of brane thicknesses
from 1.1 to 2.2.
Plots with different number of dimensions are similar to
fig.~\ref{charge}; they are always bounded above and below by the same
values, but for different values of the brane thickness.

\FIGURE{\epsfig{file=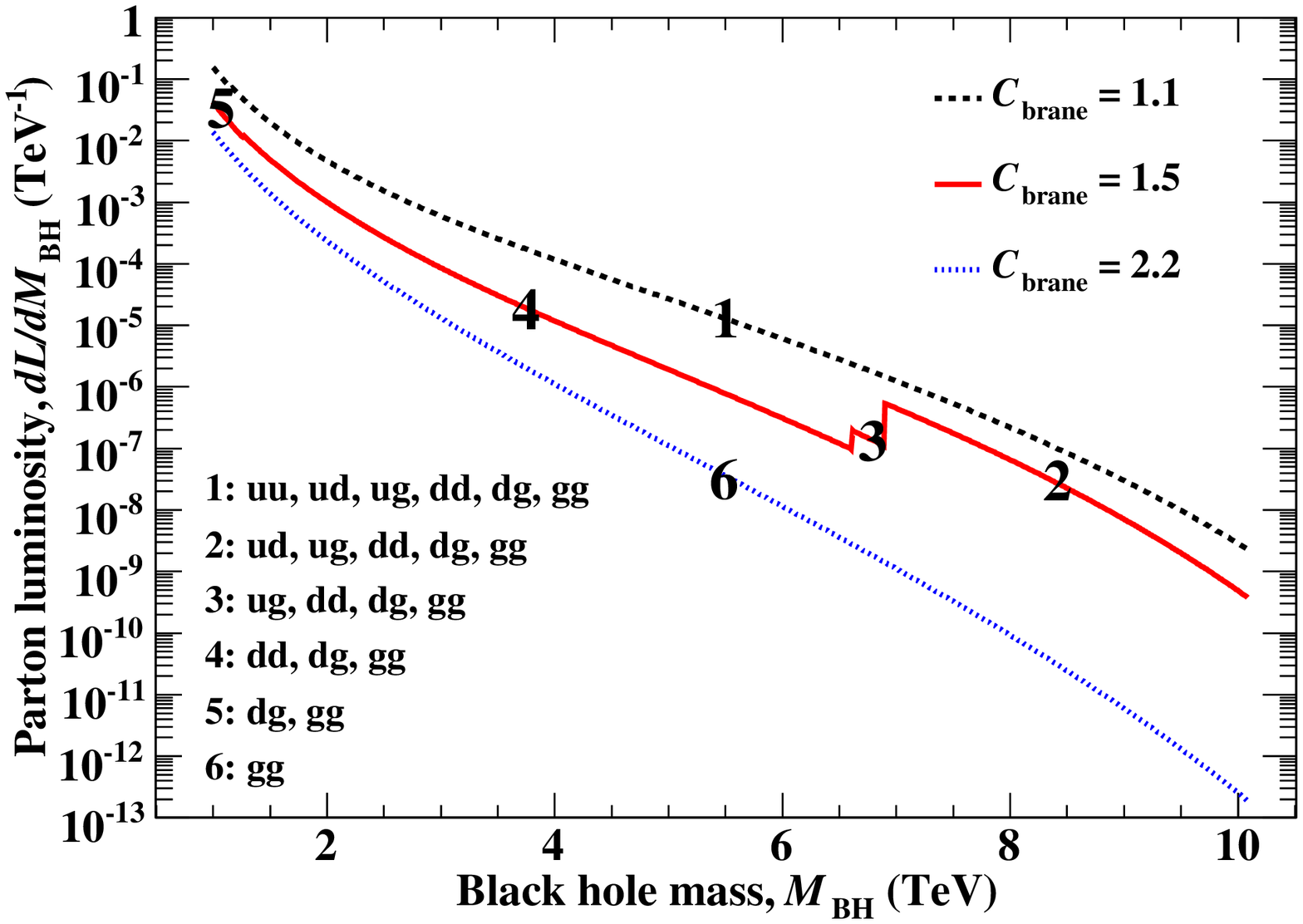,width=12cm}
\caption{Parton luminosity function versus black hole mass with the
charge condition applied for different brane thicknesses for $D=11$
and $M_D=1$~TeV.  The numbers on the plot show the different parton
contributions to the parton luminosity function.}
\label{charge}
}

For each number of dimensions, we determine the maximum brane
thickness for all partons to be included in the parton luminosity and
the minimum brane thickness for only gluons to be included in the parton
luminosity. 
The results are shown in fig.~\ref{brane}.
For a thin brane, the cross section is not affected for high dimensions.  
For a thick brane, the cross section is reduced for most number of dimensions.
For a Planck scale of 1~TeV and a brane thickness of 1~TeV$^{-1}$, the
cross section is minimal for $D\lesssim 8$, not affected for $D\gtrsim
11$, and has a range of values in the region $9\lesssim D\lesssim 10$. 

\FIGURE{\epsfig{file=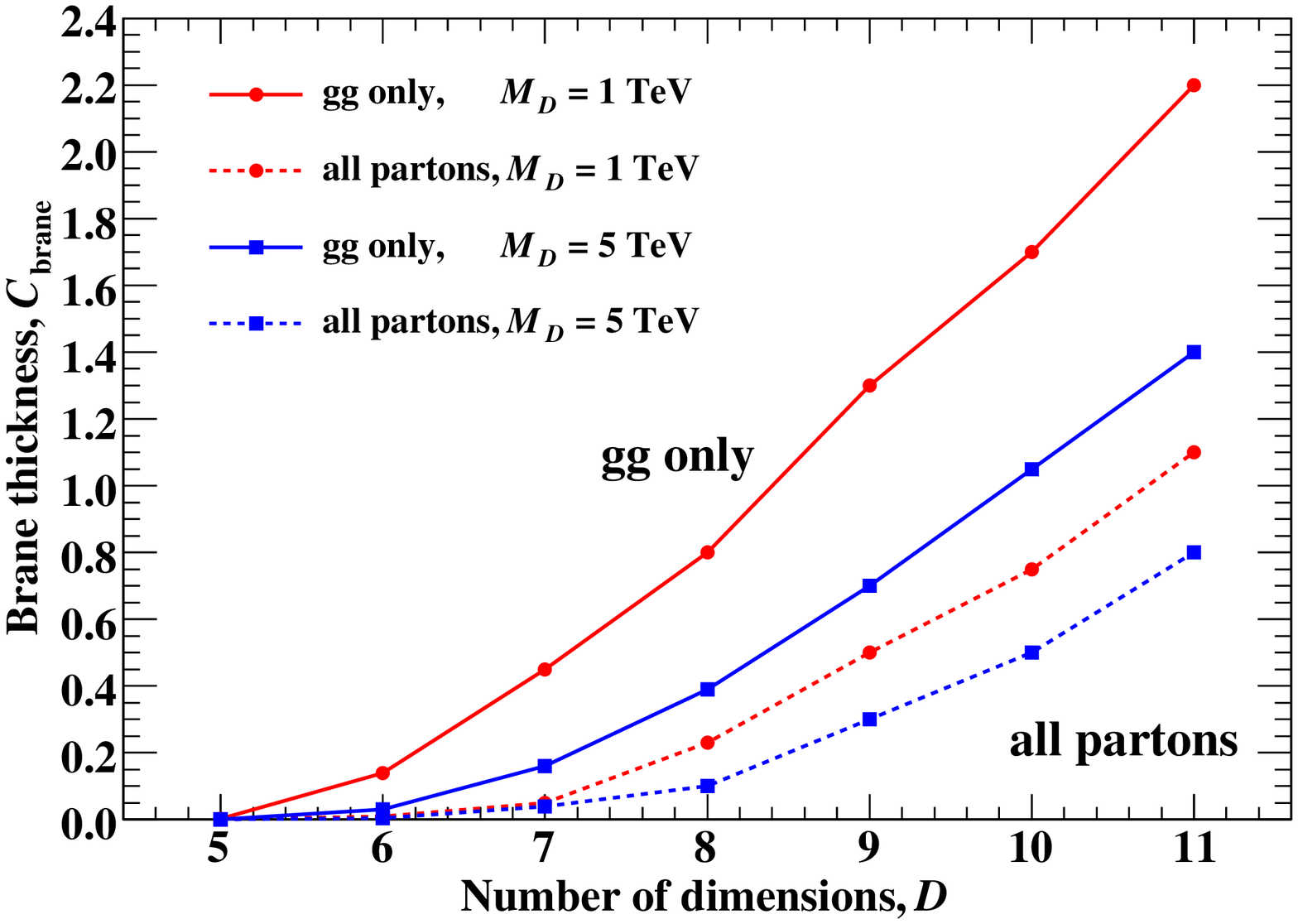,width=12cm}
\caption{Brane thickness versus number of dimensions for which all quarks
(dashed lines) and no quarks (solid lines) contribute to the cross
section.  
The circles are for $M_D=1$~TeV and the squares are for $M_D=5$~TeV.  
No quarks contribute to the cross section in the region above the solid
curves.  
The cross section is not affect by the charge condition below the region
of the dashed curves. 
Some quarks contribute to the cross section in the region between the
different curve types.}
\label{brane}
}

Using the definition of the Planck scale and electric charge, we find
that the condition for a Reissner-Nordstr{\"o}m horizon,
eq.~(\ref{eq6}), becomes 

\begin{equation} \label{eq38}
\left| C_{q_1}+C_{q_2} \right| \le \frac{M_\mathrm{BH}}{M_D}
\frac{1}{\sqrt{\alpha}\Omega_{D-2}} \sqrt{\frac{D-3}{D-2} } \left(
\frac{2\pi}{C_\mathrm{brane}} \right)^\frac{D-4}{2} \, . 
\end{equation}

\noindent
This inequality is satisfied for $D>4$ provided the brane is not too thick.
This result is contradictory to Ref.~\cite{Casadio}, where the effect of
the Standard Model confinement to the three-brane seems to have been ignored.
Table~\ref{t1} shows upper bounds on the brane thickness for the
inequality in eq.~(\ref{eq38}) to be satisfied.
However, the natural thickness of the brane in models with low-scale
quantum gravity can not be much larger than $C_\mathrm{brane} = 1$, and
thus it appears that the condition will always be satisfied for most
natural values of the brane thickness.
Our previous results indicated that a collision of two partons may not
form a black hole even though the condition for a Reissner-Nordstr{\"o}m 
horizon is satisfied.
This is because eq.~(\ref{eq38}) is a necessary, but not sufficient,
condition for black hole formation.

\TABLE{
\centering
\begin{tabular}{|l|cccccccc|} \hline
$D$                & 4 &  5  &  6  &  7  &  8  &  9  &  10 &  11 \\ \hline
$C_\mathrm{brane}$ & NA & 0.8 & 1.8 & 2.5 & 3.1 & 3.5 & 4.1 & 4.4 \\
\hline
\end{tabular}
\caption{Maximum brane thicknesses for a Reissner-Nordstr{\"o}m black hole
to form.} 
\label{t1}
}

The uncertainties in our analysis are large and particular values for
the brane thicknesses obtained in this section should be taken with
caution.   
However, the trends in values of brane thicknesses should be indicative
of a more accurate formulation of the effects of charge on the black
hole cross section.

\section{\label{sec7}Discussion}

The electromagnetic interaction between the quarks could enhance or
degrade black hole formation depending on the sign of the charges
involved.
Collisions between like-signed charged quarks will degrade black hole
formation, while collisions between opposite-signed charged quarks will
enhance black hole formation. 
It is not possible to know how effective the electromagnetic interaction
is without directly computing the subsequent temporal evolution of the
system. 

The apparent horizon analysis was carried out in a regime where QED
effects may be important and could restrict the reliability of the metric.
QED effects become important when the exterior electrostatic energy of a
point charge is equal to its rest mass.
This condition can be written as

\begin{equation} \label{eq37}
\frac{a}{r_0^{2(D-3)}}\ \lesssim\
\frac{\pi\Omega_{D-3}(2D-5)!!}{\Omega_{D-2}(2D-4)!!} \, . 
\end{equation}

\noindent
The values of the right hand side range from 0.56 to 0.68 for $4 \le D
\le 11$.
Since the apparent horizon occurs below $a = 2/(3\sqrt{3}) = 0.38$, the
condition given by eq.~(\ref{eq37}) is always satisfied.
However, the condition for the importance of QED effects is a sufficient
condition, not a necessary condition.
It is possible that QED effects are important in the neighbourhood of
the equality in eq.~(\ref{eq37}).
We can not be sure if QED effects suppress or enhance the repulsive
charge effect we have obtained.

Taking the boosted Reissner-Nordstr{\"o}m metric as a reasonable
description of ultarelativistic quarks, we have shown that charge
effects will significantly decrease the rate of black hole formation at
the LHC, if the brane is somewhat thick or if the dimensionality $D$ is
not too large. 
The charge effects can be quite large because the electromagnetic
energy-momentum tensor is proportional to $p_e^2 \sim \gamma\alpha$ and
the Lorentz factor $\gamma$ is much larger than $1/\alpha$ for
ultrarelativistic quarks.

By using parton luminosity we have not had to specify the parton-level
black hole cross section. 
As long as the cross section depends only on $M_\mathrm{BH}$, $D$, and
$M_D$, our results should be applicable for any form of the parton-level
cross section.

There remains a possibility that a black hole will form under
collision even if there is no apparent horizon on the slice we
considered, because apparent horizon formation is only a sufficient
condition for black hole formation. 
In order to specify more detailed criteria for black hole formation, it
will be necessary to study the temporal evolution of spacetime after the 
collision. 
The inclusion of the spin of the incoming particles is also required. 
Inclusion of QED effects and brane effects on gauge-field confinement
may also be necessary. 

\section*{Acknowledgments}

I thank Hirotaka Yoshino for helpful discussions.
This work was supported in part by the Natural Sciences and Engineering
Research Council of Canada.


\end{document}